\documentclass[fleqn,10pt]{wlscirep}
\usepackage[utf8]{inputenc}
\usepackage[T1]{fontenc}
\title{Three ways to select from two attosecond pulses}

\author[1,*]{Katalin Kov\'acs}
\author[1]{Valer Tosa}
\affil[1]{National Institute for R\&D of Isotopic and Molecular Technologies, RO-400293 Cluj-Napoca,
Romania}

\affil[*]{kkovacs@itim-cj.ro}

\begin{abstract}
We demonstrate that by filtering high harmonics it is possible to obtain two different single attosecond pulses (SAP) resulting from naturally separated spectral domains formed during propagation in the macroscopic medium. We propose a feasible experimental configuration in which one can obtain a SAP in a lower energy domain (<300 eV), or another SAP in a higher energy domain (>300 eV). Without filtering, a double attosecond pulse emission with fixed temporal separation is obtained. The gap between the two spectral domains is close to the onset of the water window.
\end{abstract}
\begin{document}

\flushbottom
\maketitle
%
%
\thispagestyle{empty}

\section*{Introduction}

High-order harmonic generation (HHG) with near-infrared (NIR) and mid-infrared (MIR) driving laser pulses is the well-established table-top method to obtain highly coherent pulses in the extreme ultraviolet (XUV) and soft X-ray spectral domain with temporal width in the order of tens-to-hundreds of attoseconds \cite{Li2017, Gaumnitz2017}, paving the way for the development of Attosecond Science \cite{Villeneuve2018, Krausz2016}.
Applications of the attosecond pulses range from studying the real-time electron dynamics in atoms and molecules, to strong-field induced processes in solids or biomolecular imaging \cite{Krausz2014, Calegari2016, Nisoli2017, Schultze2014, Helk2019}.
For the applications it is usually required that the attosecond pulses should be bright, reach high energy like the water window (282-533 eV), and in many cases it should be only one isolated attosecond pulse \cite{Sansone2006}. One important research challenge is to elaborate methods to control the generation and characteristics of attosecond pulses \cite{Winterfeldt2008}.

According to the classical cutoff-law $E_{cut}=I_p+3.17\cdot U_p$ where $U_p \sim E^2\cdot\lambda^2$ is the ponderomotive energy gained by the electron in its excursion outside the atom. MIR driving pulses are advantageous because the available photon energy can be extended to the water-window. However, due to the wave packet spreading in the continuum the efficiency of HHG scales as $\lambda^{-(5-6)}$ \cite{Tate2007}. HHG by long wavelength pulses has gained attention since the laser technology has evolved and ultrashort, CEP stable laser pulses at high intensity are available also in the MIR spectral domain \cite{Gaumnitz2017, Fu2018, Negro2018, Cardin2018, Ciriolo2017, Kuhn2017, Yin2016, Darginavicius2012, Ishii2012, Vozzi2007}.

As known, with few-cycle driving pulses, there exists the possibility to obtain experimentally one single attosecond pulse (SAP) \cite{Sansone2006}. However, whenever in an experiment a quasi-continuous high-harmonic spectrum is measured, a basic question arises: is this the signature of a SAP, or the quasi-continuous spectrum is the superposition of different radial components emitted shifted in time? When the driving pulse is very short (few-cycle) and very intense, ionization gating \cite{Ferrari2010} may occur, and a SAP is obtained. In the same time the high-harmonic spectrum strongly depends on the carrier-to-envelop phase (CEP) of the driving pulse \cite{Sola2006, Sansone2009}: with changing CEP the harmonic spectrum changes from continuous to modulated and indicates the change from one single emission to interference between two successive emissions. Quantum path analysis reveals that the short and long trajectory components are influenced differently by the change in the driving field's CEP \cite{Sansone2004, Sansone2004a}.

In this work we demonstrate that the total macroscopic spectrum can be generated in two distinct spectral regions: a low energy one corresponding to short trajectory emission in the plateau region and a high energy one corresponding to cutoff trajectories. We show that these two regions correspond to two separated bursts in time, so that by spectral filtering one can select the burst of low energy, the other burst of high energy, or keep both bursts with fixed delay. We show that by varying the CEP we can vary the relative intensity of the two bursts, and we identify the experimental parameters that act like "control knobs". The configuration we investigate is not optimized for SAP generation, but rather we concentrate on the temporal and spectral separation of the double pulse structure.

\section*{Modeling tools and proposed experimental configuration}

\subsection*{Model}

The numerical simulations have been carried out using the extended version of the 3D non-adiabatic model presented in \cite{Tosa2003}. In this framework we first solve the pulse propagation in ionized gas medium and then calculate the atomic dipole induced by the driving field using the strong-field approximation \cite{Lewenstein1994}. The  macroscopic harmonic signal is the result of the coherent superposition of the single dipole emissions in each spatial point. The interaction region has cylindrical symmetry in the simulations.

1. The propagation of the fundamental laser pulse in a gaseous medium is very important to be accurately treated \cite{Gaarde2008}. Since the model has been described in several papers \cite{Tosa2003, Tosa2005a, Tosa2005b, Tosa2009}, we mention here only the expression for the space-time dependent refractive index of the medium:
\begin{equation}
    \eta_{eff}(r,z,t)=\eta_1(r,z,t)+\eta_2 I(r,z,t)-\frac{\omega_p^2(r,z,t)}{2\omega^2},
    \label{eta}
\end{equation}
where the first linear term $\eta_1=1+\delta_1-i\beta_1$ contains the dispersion and absorption of neutrals; the second term accounts for the optical Kerr effect dependent on the nonlinear refractive index ($\eta_2$) and laser intensity ($I(r,z,t)$); the last term expresses the plasma dispersion, with plasma frequency $\omega_p=(4\pi e^2 n_e/m)^{1/2}$, $n_e$ is the free electron number density, $e$, $m$ are the electron charge and mass, $\omega$ is the driving pulse's central frequency. For the calculation of the ionization rate of He we use the model from \cite{Scrinzi1999}. The non-trivial spatial and temporal variation of the refractive index induces the reshaping of the driving pulse.
The ionization front of a strong laser pulse induces rapid variation of the medium's refractive index, due to this the pulse itself suffers distortions in its spatial shape as well as in its temporal and spectral properties \cite{Tosa2005a,Tosa2005b}.

2. The single dipole will be calculated in each spatial point using the {\it propagated} fundamental pulse, therefore the harmonic dipole radiation inherently contains the properties of the modified fundamental field. The radial variation of the driving pulse implies that all intensity-dependent terms will be non-homogeneous radially. This radial non-homogeneity of the propagated fundamental pulse will be reflected in the whole process of HHG.

3. The nonlinear dipole radiations are sources for the macroscopic harmonic field which builds up propagating in the same gas medium partially ionized by the fundamental laser pulse. Absorption and dispersion of the harmonic field are taken into account. 

4. The simulation method has been extended to include specific experimental conditions, and to better elucidate the underlying physics.  Quantum trajectory analysis \cite{Kovacs2010b} was used as an independent verification of the temporal and phase properties of attosecond pulses at sub-optical-cycle temporal resolution. The method is based on numerically solving the complex-valued saddle-point equations \cite {Kopold2000, Milosevic2006} in the case of arbitrary pulse shape, for example the propagated and distorted driving pulse for the present study. Finally, the Hankel transform was used to transport the harmonic field form the medium exit to the far-field.

\subsection*{Proposed experimental configuration}
\label{config}

The main goal of this paper is to demonstrate that in an experimentally feasible configuration one can obtain two successive attosecond pulses which are generated by two naturally separated spectral regions. For this purpose we constructed a case with experimental parameters similar to those reported in Ref. \cite{Cardin2018}. The driving wavelength of $\lambda = 1700$ nm has the advantage that it allows for extended single-dipole cutoff; the extremely short duration of 6 fs ensures that we will have only few emission events per pulse; the 9 mm long gas medium contains He at 200 Torr pressure, and is placed in the converging beam, such that it ends in the nominal laser focus position ($f=50$ cm).

In order to verify the robustness of the spectral separation, and which parameters influence its existence, we varied the carrier-envelop phase (CEP) of the driving pulse (0, $\pi/8$, $-5\pi/8$) and the input laser pulse energy i.e. the peak intensity in the input plane ($9\cdot 10^{14}$ and $7\cdot 10^{14}$ W/cm$^2$).

\section*{Results and discussion}
\subsection*{Driving pulse propagation and harmonic spectrum}

\begin{figure}[htbp]
\includegraphics[width=18cm]{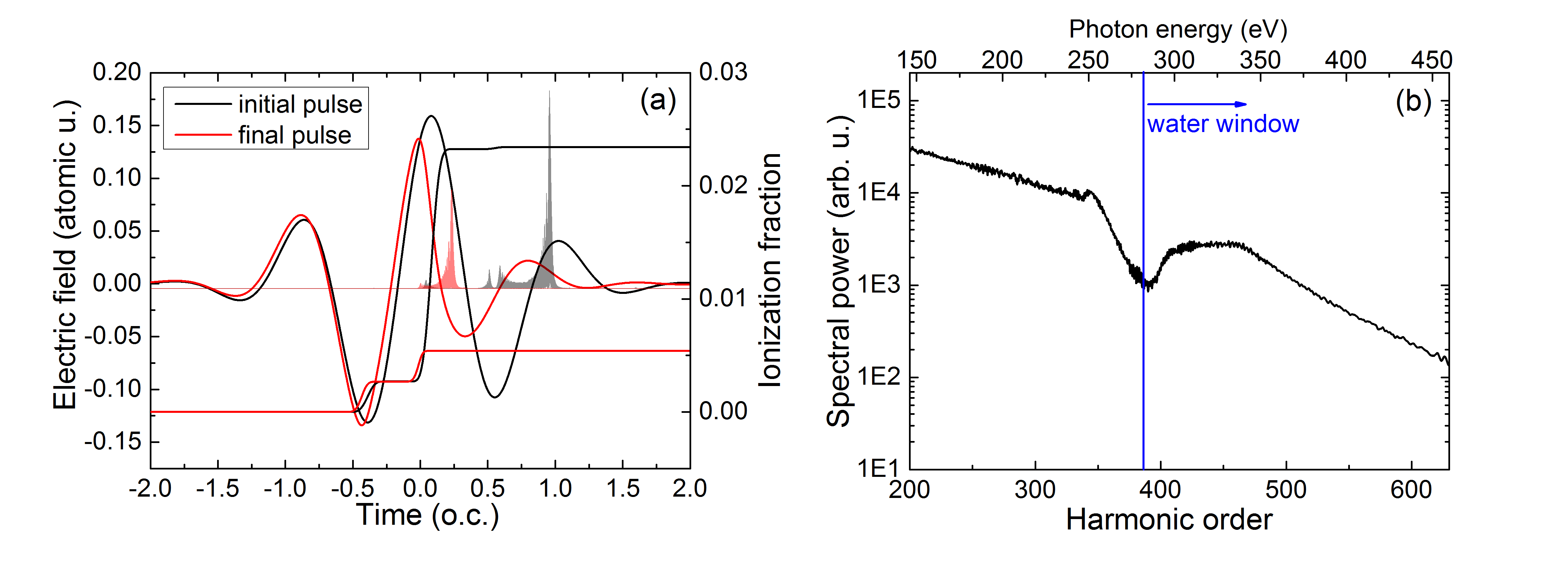}
\caption{(a) Driving pulse shape along with the temporal ionization dynamics at the beginning (black) and at the end (red) of the propagation in 200 Torr He. Shaded in the background we show the dipole radiation filtered between harmonic orders H200 -- H630. Time is expressed throughout the paper in units of the driving field's optical cycle ($T$), the nominal pulse peak is time zero. (b) Radially integrated harmonic power spectrum at the end of the interaction region. The blue vertical line indicates the beginning of the water window spectral domain at 282 eV.}
\label{pulse}
\end{figure}

In Fig. \ref{pulse}(a) we show the temporal shape of the driving laser pulse at the beginning and at the end of propagation with black and red lines, respectively, for the case: $I_{peak} = 9\cdot 10^{14}$ W/cm$^2$; CEP = 0; medium length = 9 cm. The ionization dynamics during the pulse is also shown, with black (initial) and red (final) lines. The well known self-phase-modulation and the decrease of the peak intensity are clearly present due to the steep ionization front around the peak of the pulse. This effect scales with $\lambda^2$, therefore longer wavelength pulses are particularly affected even by a relatively low ionization level, which in this particular case is between 2.3\% and 0.5\%. The attosecond pulse emissions shown shaded in the background indicate that two main events take place during the laser pulse: (1) the earlier ($T\approx 0.2$) burst is due to the recombination of electrons ionized around $T=-0.3$; this emission can contain all spectral components up to the cutoff. (2) The second burst at $T\approx 0.6$ is due to recombining electrons which were ionized right after the pulse peak. 

Fig. \ref{pulse}(b) shows the radially integrated harmonic power spectrum at the exit from the interaction region, between harmonic orders 200 and 630 ({\bf H200 -- H630}) which corresponds to {\bf 146 -- 460 eV}. After 250 eV we observe a gap in the spectral power, or in other words we have a double plateau structure: a first cutoff around 255 eV, then a revival at higher photon energies 300 -- 365 eV. The gap in the spectrum is at the beginning of the water window region which could be essential for applications in soft x-ray microscopy. 

Since the spectrum in Fig. \ref{pulse}(b) is a volume integrated quantity it is essential to examine the radial structure of the harmonic field at the medium exit. In the next subsection we will explore the radial structure of the harmonic fields, the main goal being to demonstrate the spectral separation of the two attosecond pulses, as well as to identify the "control knobs" which switch on/off this spectral separation.

\subsection*{Spectrally separated attosecond pulses}

\begin{figure}[htbp]
\includegraphics[width=18cm]{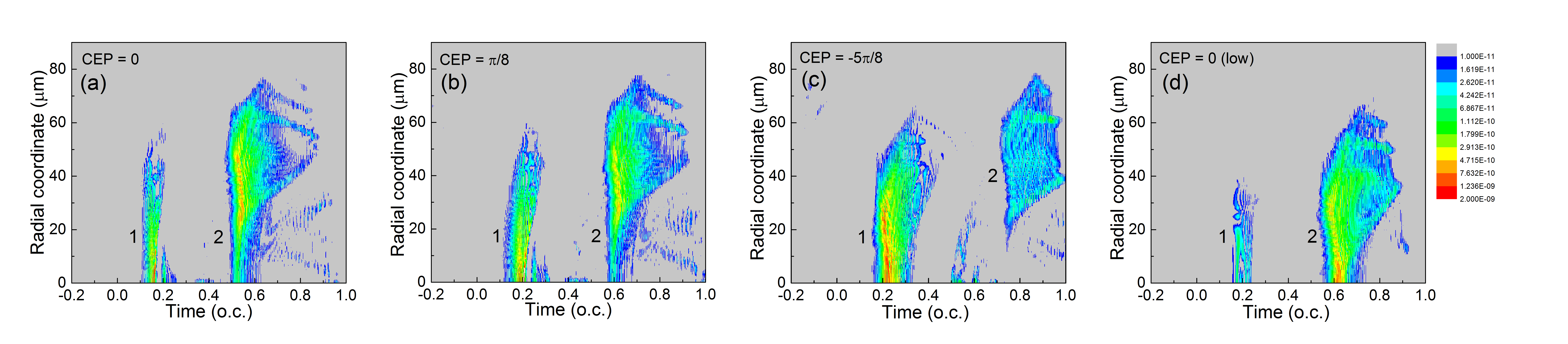}
\caption{Temporal--radial maps of the attosecond pulses in the H200 -- H630 spectral domain. Snapshots are taken at the exit of the interaction region. Color scale spans two orders of magnitude, logarithmic scale, arbitrary units. In (a)-(c) panels the input laser intensity is $9\cdot 10^{14}$ W/cm$^2$, and the CEP is 0, $\pi/8$ and $-5\pi/8$, respectively. In panel (d) the input intensity is $7\cdot 10^{14}$ W/cm$^2$ and CEP = 0. }
\label{tr}
\end{figure}

In Fig. \ref{tr} we show the temporal--radial maps of the attosecond pulses filtered in the H200 -- H630 spectral domain. The snapshots are taken at the exit of the interaction region, which is 9 mm long in all cases. The parameters that are modified in successive panels are specified.  With reference to the single-dipole emissions shown in Fig. \ref{pulse}(a) we observe that the double pulse structure has been maintained during propagation, but the long trajectory components of the second emission are not present in the final macroscopic harmonic field as they did not find good phase matching conditions during propagation. For future reference we labelled the earlier and later emissions with 1 and 2, respectively.

\begin{figure}[htbp]
\includegraphics[width=18cm]{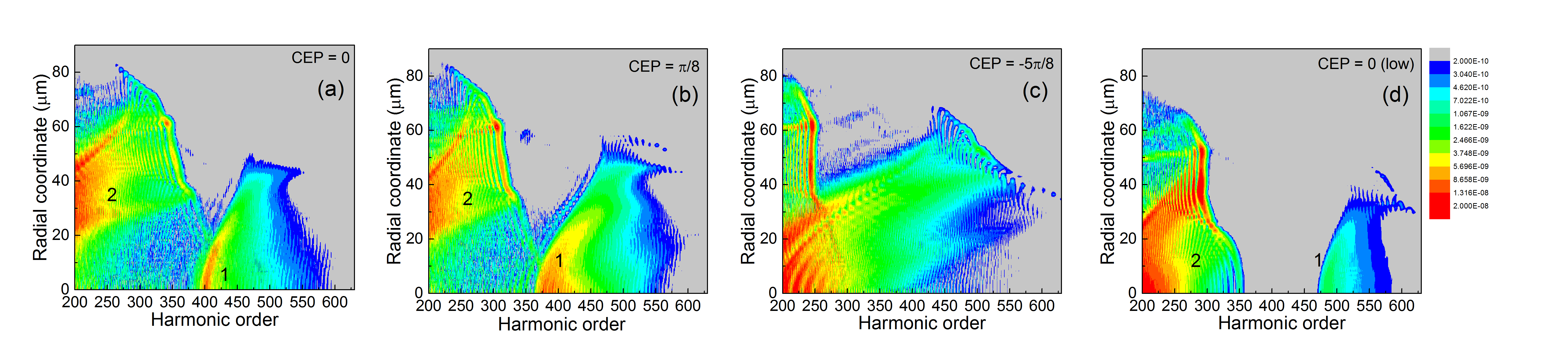}
\caption{Radial maps of the harmonic radiation at the exit of the interaction region. The four cases are identical to those from Fig. \ref{tr}. Snapshots are taken at the exit of the interaction region. Color scale spans two orders of magnitude, logarithmic scale, arbitrary units.}
\label{wr}
\end{figure}

In Fig. \ref{wr} we show the radial map of harmonics in the H200 -- H630 spectral region, also at the exit of the medium, in the same four cases as presented in Fig. \ref{tr}. The temporal--radial and the spectral--radial maps, are complementary to each other, they are two projections of a three-dimensional space with temporal, spectral and radial coordinates. Therefore, the information contained in Figs. \ref{tr} and \ref{wr} have to be interpreted in pairs. 

In Fig. \ref{tr}(a) we observe that emission 1 is developed close to the axis, while emission 2 has the main contribution off-axis, in a ring of $r\in [20,50]$ $\mu m$. Comparing this feature with Fig. \ref{wr}(a), we can clearly identify that emission 1 is associated with higher orders having the maximum at H400 -- H420. Emission 2 contains lower orders up to H350, maximum being around H200 -- H250. The spectral separation of the attosecond pulses 1 and 2 is visible in Fig. \ref{wr}(a), the gap is around H380 (277 eV). The maps are qualitatively similar for a slightly modified CEP, as shown in Figs. \ref{tr}(b) and \ref{wr}(b) where CEP = $\pi/8$. The spectral separation of the two attosecond pulses still persists, but the gap between the frequency domains moved to lower orders, namely around H350 (255 eV). When we modified the driving pulse's CEP to $-5\pi/8$, as seen in Fig. \ref{wr}(c), the spectral separation does not happen, emissions 1 and 2 labeled in Fig. \ref{tr}(c) cannot be identified in the corresponding frequency map in Fig. \ref{wr}(c). Further, if we reduce the input laser intensity to $7\cdot 10^{14}$ W/cm$^2$, and keep the initial CEP = 0, the spectral separation is very clear, with a generous spectral gap in the H350 -- H450 (255-330 eV) spectral region, as seen in Fig. \ref{wr}(d). As expected, however, the intensity of the high-energy emission 1 is reduced compared to the previous cases. 

The results so far indicate that (1) the spectral separation is CEP-dependent; (2) the spectral gap can be slightly controlled by the CEP of the driving pulse, but through modifying the laser pulse intensity in the input plane we can influence the spectral position of the gap, as well as the bandwidth of this gap. (3) This tunability can be explored in order to tune the spectral separation according to available XUV filters. For example Sn/SiN and In/SiN filters work for 3-5 nm wavelengths (248-413 eV).

\begin{figure}[htbp]
\includegraphics[width=18cm]{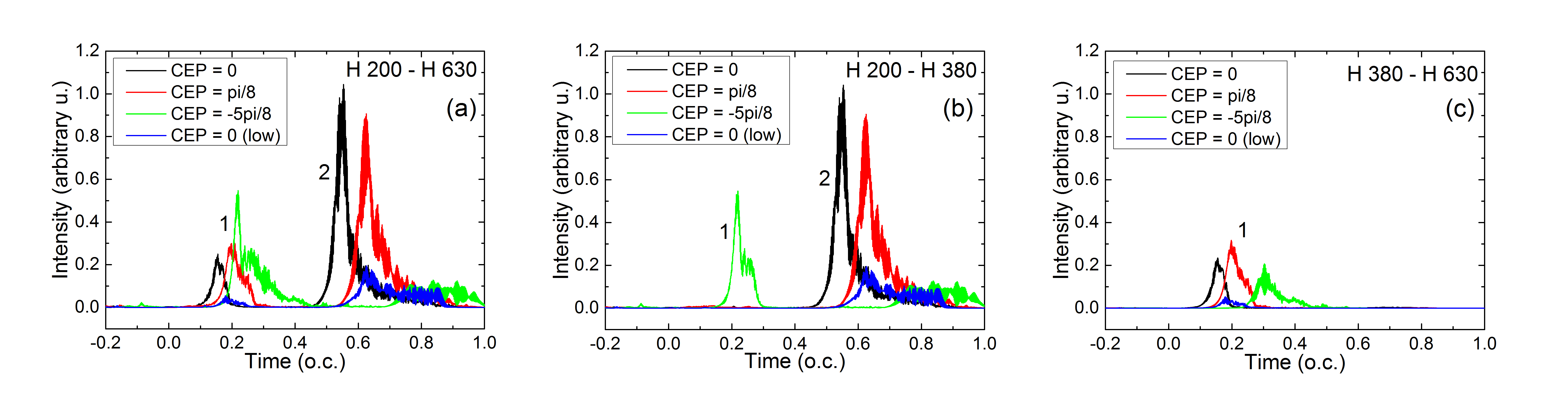}
\caption{Radially integrated attosecond pulses in the far-field. (a) Full spectrum between H200 -- H630. (b) The lower part of the spectrum between H200 -- H380. (c) The high part of the spectrum between H380 -- H630. In each panel the colors identify the four cases: (black) CEP = 0; (red) CEP = $\pi/8$; (green) CEP = $-5\pi/8$; (blue) CEP = 0 (low) meaning $I_{peak }=7\cdot 10^{14}$ W/cm$^2$.}
\label{fart}
\end{figure}

The combined information contained in the panels of Fig. \ref{tr} and Fig. \ref{wr} is helpful but we need evidence of the temporal AND spectral separation of the attosecond pulses, simultaneously. In simpler terms: we need to verify that the emission 1 contains only high-order spectral components, while emission 2 is composed only by lower order harmonics. The final confirmation is given in Fig. \ref{fart}, where we show for the four cases the radially integrated attosecond pulses in the far-field, placed 1 m in vacuum from the medium, assuming no focusing. 

When the full spectrum is inverse Fourier transformed, the double attosecond pulse structure is visible for all parameter combinations included in this paper, see Fig. \ref{fart}(a). The properties of the double pulse for the cases $I_{peak }=9\cdot 10^{14}$ W/cm$^2$ with CEP = 0 and CEP = $\pi/8$ are qualitatively similar, there is a fixed delay of $\tau=0.4$ T (2.26 fs) between emissions 1 and 2. The intensity ratio $I_1/I_2$ of emission 1 to emission 2 are: $I_1/I_2=0.24$ for CEP = 0, and  $I_1/I_2=0.33$ for CEP = $\pi/8$. 

When a low-pass spectral filter is applied such to keep only radiation below H380 (277 eV), as seen in Fig. \ref{fart}(b) emission 2 is conserved with full intensity in cases with CEP = 0 and CEP = $\pi/8$, and a strong single attosecond pulse is obtained. The CEP = $-5\pi/8$ keeps the temporal separation, however, both emissions 1 and 2 contain the low order components. This result is in perfect agreement with the spectral -- radial map in Fig. \ref{wr}(c). 

Further, if our filter is high-pass and we keep only the high-energy part of the spectrum with harmonics above H380 (>277 eV) a single attosecond pulse from emission 1 is generated  in all four cases, as shown in Fig. \ref{fart}(c).

\subsection*{Mechanism of spectral separation}

Next, we want to explain the spectral separation of the two emissions. First, it is obvious that the spectral components in emission 2 cannot contain very high-order harmonics due to the reduced average intensity of the driving pulse in the trailing edge. The question remains: what causes that in emission 1 only the high energy components are present after propagation?  Indeed, it is not evident why for certain CEP values the lower orders are absent from emission 1.

\begin{figure}[htbp]
\includegraphics[width=18cm]{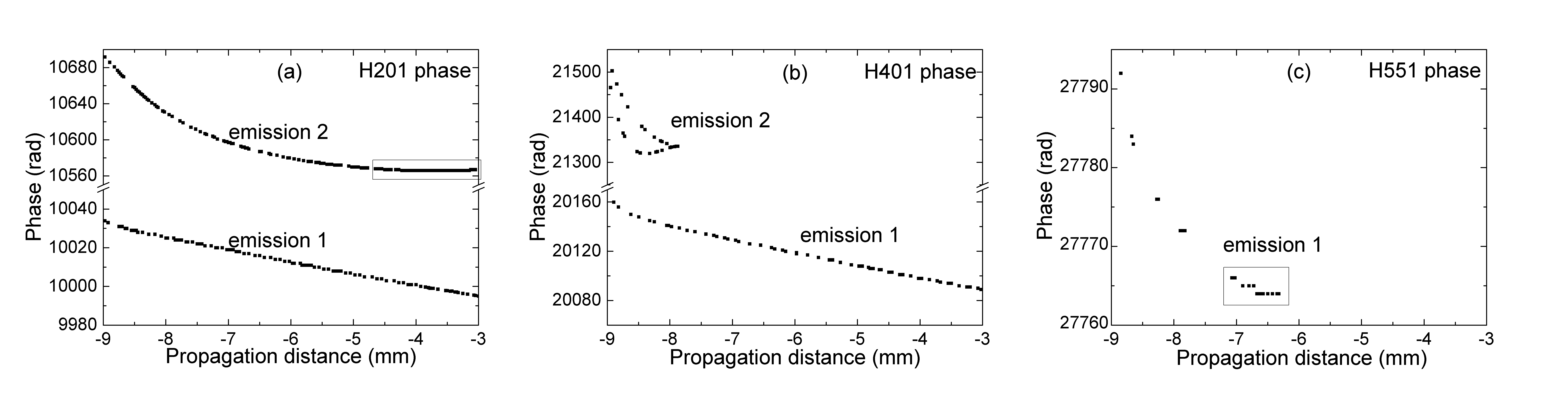}
\caption{(a) Phase of H201 as it varies along propagation in emissions 1 and 2. The rectangle emphasizes the section along propagation where the harmonic phase is constant. (b) Phase of H401 as it varies along propagation in emissions 1 and 2. The lack of constant phase zone hinders the coherent construction of H401. (c) Phase of H551 as it varies along propagation in emission 1. The rectangle emphasizes the section along propagation where the harmonic phase has $<\pi$ rad variation.}
\label{traject}
\end{figure}

Trajectory analysis gives the answer for this question. We studied in detail the case $I_{peak }=9\cdot 10^{14}$ W/cm$^2$, CEP = 0, and calculated the phases of trajectories responsible for harmonics H201 (present only in emission 2), H401 (in the spectral gap) and H551 (present only in emission 1), as presented in Fig. \ref{traject}. The results emphasize the crucial role of the propagation effects which determine the final macroscopic harmonic signal. H201 is clearly present at dipole level in both emissions 1 and 2, which is confirmed by the quantum trajectory calculations and shown in Fig. \ref{traject}(a). However, in emission 1 the phase of this harmonic varies significantly along propagation, $\approx$ 40 rad during the 9 mm medium. In emission 2 the trajectory responsible for H201 benefited from constant phase in the $z\in [-5; -3]$ mm section, and therefore could build up constructively. Harmonics in the spectral gap, in particular H401 examined by us (Fig. \ref{traject}(b)), have strong phase variation all along propagation, making it impossible to grow in intensity. On the other hand, harmonic H551 (Fig. \ref{traject}(c)) in emission 1 had $<\pi$ phase variation in the $z\in [-7; -6]$ mm portion and in this constant phase section the radiation could build up constructively. Trajectory analysis confirms again, that important aspects of the HHG process are basically influenced by propagation effects, therefore macroscopic modeling of HHG is essential.

\section*{Conclusions}

In conclusion, we numerically demonstrate that high harmonics generated in He gas by few cycle MIR pulses are emitted in two separated spectral domains formed during propagation in a long gas medium. We also show that these two spectral domains are emitted in two separate pulses so that by filtering it is possible to select one of them or both.

From the fundamental point of view, we emphasized the importance of the macroscopic propagation effects in shaping the final XUV emission. The phenomenon described here as spectral separation of successive attosecond pulses can be an example of space-time coupling in nonlinear optics, good candidate to be further explored and exploited experimentally.

From the practical point of view, we proposed a HHG configuration based on a feasible experimental setup with the main advantage that it allows the generation of two attosecond pulses having separate spectral content, but both in the XUV regime. In this configuration one can keep the total flux of one emission, knowing the low efficiency of the HHG process especially with increasing driving wavelength \cite{Tate2007}.

Further investigations are ongoing to explore such generation schemes using NIR pulses and/or higher pulse energies. 

\bibliography{Optica2019_references}

\section*{Acknowledgements}

Financial support is acknowledged for the Romanian National Authority for Scientific Research and Innovation project RO-CERN 03ELI (PROPW), UEFISCDI public institution under the Romanian Ministry of Education project PN-III-P4-ID-PCE-2016-0208.
We acknowledge the access to the Data Center at INCDTIM Cluj-Napoca.






\end{document}